\documentclass[twocolumn,preprintnumbers,amsmath,amssymb,floatfix]{revtex4}
\usepackage{graphicx}% Include figure files
\usepackage{dcolumn}% Align table columns on decimal point
\usepackage{bm}% bold math

\begin{document}

%\preprint{APS/123-QED}

\title{Far Field Monitoring of Rogue Nuclear Activity with \\
an Array of Large anti-neutrino Detectors}

\author{Eugene H. Guillian}
%\altaffiliation[Also at ]{Department of Physics, University of Hawaii,
%Manoa}%Lines break automatically or can be forced with \\
%\author{Second Author}%
\email{guillian@owl.phy.queensu.ca}
\affiliation{%
Department of Physics, University of Hawaii, Manoa, 2505 Correa Rd.,
Honolulu, HI 96822, USA\footnote{Current Address: Department of Physics, Queen's University, Stirling Hall, Kingston, ON, K7L 3N6, Canada}
%This line break forced with \textbackslash\textbackslash
}%

%\author{Charlie Author}
% \homepage{http://www.Second.institution.edu/~Charlie.Author}
%\affiliation{
%Second institution and/or address\\
%This line break forced% with \\
%}%

\date{\today}% It is always \today, today,
             %  but any date may be explicitly specified

\begin{abstract}

The result of a study on the use of an array of large anti-neutrino
detectors for the purpose of monitoring rogue nuclear activity is
presented.  Targeted regional monitoring of a nation bordering large bodies of water with no pre-existing legal nuclear activity may be possible at a cost of about several billion dollars, assuming several as-yet-untested schemes pan out in the next two decades.  These are: (1) the enabling of a water-based detector to detect reactor anti-neutrinos by doping with GdCl$_3$; (2) the deployment of a KamLAND-like detector in a deep-sea environment; and (3) the scaling of a Super-Kamiokande-like detector to a size of one or more megatons.  The first may well prove feasible, and should be tested by phase-III Super-Kamiokande in the next few years.  The second is more of a challenge, but may well be tested by the Hanohano collaboration in the coming decade.  The third is perhaps the least certain, with no schedule for construction of any such device in the foreseeable future.  In addition to the regional monitoring scheme, several global, untargeted monitoring schemes were considered.  All schemes were found to fail benchmark sensitivity levels by a wide margin, and to cost at least several trillion dollars.
\end{abstract}

%\pacs{Valid PACS appear here}% PACS, the Physics and Astronomy
                             % Classification Scheme.
%\keywords{Suggested keywords}%Use showkeys class option if keyword
                              %display desired
\maketitle

\section{Introduction}

The human race first tapped into nuclear energy with the success of
the Manhattan project.  Ever since, the practical know-how regarding
the use of this source of energy has expanded and spread, and, so far
as civilization as we know it continues to exist, this, no doubt, will
continue to be the case.  The spread of practical know-how in this
area, however, presents a threat to peace, since there will always be
desperate characters among the world's political leaders, and it is a
matter of time before one such leader gets access to this know-how and
decides to use it indiscriminately against his enemies.

Monitoring regimes exist to guard against the uncontrolled spread of
nuclear technology and the detonation of nuclear bombs.  The
International Atomic Energy Agency (IAEA) works under the auspices of
the United Nations to make sure that nations that use nuclear energy
do so only for peaceful purposes~\cite{iaea}.  Another monitoring regime is the
Comprehensive Test Ban Treaty (CTBT), which is an agreement among
nations to ban all nuclear explosions~\cite{ctbt}.  As recent world events (the detonation of a fission bomb by Pakistan in 1998, and the current political crisis involving nuclear activities in North Korea and Iran) have
shown, however, neither regime has proved sufficient to curb the
spread of nuclear technology nor the detonation of bombs.  Clearly,
the flaws in the regimes are mostly political.  For instance, the
detonation of nuclear bombs by Pakistan in 1998 was not against the
CTBT because Pakistan is not a signatory.  Also, the recent events in
North Korea and Iran have little to do with monitoring techniques,
but, rather, with flaws in the political process that allows
headstrong political leaders to use nuclear threats as political
bargaining chips.

Although much of the problems with today's monitoring regime is
political, some of the political problems are abetted by
insufficiencies in the monitoring techniques.  For instance, in 2002,
after mounting tensions with the United States and her allies, North
Korea expelled United Nations inspectors and threatened to restart its
nuclear facilities in Yongbyon~\cite{cnn_north_korea_timeline}.  Once the inspectors were ousted, it
was impossible to tell whether or not the North Koreans had actually
carried through with their threat to reprocess nuclear fuel.  This
scenario is made possible by the fact that the IAEA monitoring
technique requires the cooperation of participants.  Clearly, a more
robust monitoring regime requires far-field monitoring techniques that
do not depend on participant cooperation.

Such techniques are already in use to monitor nuclear explosions by
the CTBT (seismology, hydroacoustics, infrasound, and radionuclide
monitoring)~\cite{ctbt}, but they are useless for detecting nuclear reactor
operation because a reactor burns nuclear fuel at a steady rate, and it does not release redionuclides into the environment.
Far-field monitoring, however, is possible in principle using anti-neutrinos
produced in nuclear fission.  Indeed, the KamLAND experiment~\cite{kamland} detects
anti-neutrinos from nuclear reactors at an average distance of about
180 km.  anti-neutrinos are electrically neutral particles produced in
nuclear fission; they interact with matter only via the {\it weak
nuclear force}.  Because of this, anti-neutrinos can easily travel through
hundreds of kilometer of matter with almost no probability of
interaction with the intervening material.  This feature of the anti-neutrino makes its detection very difficult; however, given a big enough target, a sufficiently long exposure time, and a sufficiently low background level, they can be reliably detected.

The purpose of the study presented herein is to determine the
feasibility of using anti-neutrino detection for far-field monitoring
of both nuclear reactor operation and fission bomb detonations.  At
the most basic level, the feasibility of this technique has been
established by KamLAND.  However, they were helped by the very large
signal due to the unusually large concentration of nuclear reactors in Japan\footnote{As of 2002, Japan has 16 nuclear reactor plants producing a total power of 130~GW$_{\mbox{\footnotesize th}}$~\cite{rmp}.}.
In a realistic far-field monitoring scenario, the signal is expected
to be tiny -- probably much less than 100~MW$_{\mbox{\footnotesize th}}$ (a typical
commercial nuclear reactor power is about 2500~MW$_{\mbox{\footnotesize th}}$).  We
have found that a regionally targeted monitoring regime -- e.g. the
monitoring of nuclear reactor operations in North Korea -- is
may be possible at a projected cost of several billion dollars, as long as several as-yet-untested schemes pan out in the coming decade.  We also considered the possibility of setting up a global array
of large anti-neutrino detectors to detect surreptitious nuclear
fission activity anywhere in the world.  This was found to miss benchmark sensitivity levels by a wide margin, and to be
unrealistic because of the prohibitive projected cost on the order of trillions
of dollars.

\section{anti-neutrinos Produced in Nuclear Fission}

Nuclear reactors and fission bombs make use of the energy released by
splitting heavy nuclei (primarily uranium and plutonium).  The former are
designed to keep the rate of splitting constant so that energy is
released at a steady rate, whereas the latter is designed to cause the
energy to be released in a very short period.  In both cases, the
daughter nuclei from the splitting of uranium and plutonium are
unstable and undergo radioactive decay; an anti-neutrino is produced
from every beta decay.

The rate of anti-neutrino production in a nuclear reactor is directly
proportional to its thermal power.  Each nuclear fission releases
about 200~MeV (million electron volts) of thermal energy\footnote{One
electron volt is the amount of energy that an electron gains when it
moves from the ``$-$" to ``+" terminal of a 1-volt battery.  The energy
released by nuclear fission, therefore, is hundreds of million times
as large as the energy released by a typical battery used in every-day
life}, which is equal to $3.2 \times 10^{-11}$ Joules.  A typical
nuclear reactor has a thermal power of one gigawatt, or $10^9$ Joules
per second.  The number of fissions per second required to produce
this power, therefore, is $10^9$~J/s divided by $3.2 \times
10^{-11}$~J, which is equal to $3.1 \times 10^{19}$.  Finally, since
about 6 anti-neutrinos are produced per fission, we find that $1.9
\times 10^{20}$ anti-neutrinos per second are produced for every one
gigawatt of thermal power.

The corresponding calculation for a fission bomb proceeds similarly.
The strength of a fission bomb is usually quoted in terms of its
``yield", which is the mass of TNT that produces the same amount of
energy.  A small fission bomb has an yield of about 1000 ton of TNT,
and one metric ton of TNT releases $4.18 \times 10^{9}$ Joules, so
this bomb releases $4.18 \times 10^{12}$ Joules of energy.  As in the
case of a nuclear reactor, each fission releases 200~MeV of energy,
and 6 anti-neutrinos are produced per fission.  Thus $1.30 \times
10^{23}$ anti-neutrinos are released for every kilo-ton of fission bomb
yield.  Unlike a reactor, which releases anti-neutrinos at a steady
rate, a fission bomb releases the anti-neutrino impulsively over a
period of several seconds; almost no anti-neutrinos are emitted after
about 10 seconds from the blast~\cite{adam_bernstein}.

\section{Detecting anti-neutrinos}

The stuff that composes the material world is responsive to the
electromagnetic force.  It is this force that keeps a ball from going
through one's hand when caught.  anti-neutrinos, unlike ordinary stuff,
are unresponsive to the electromagnetic force.  Consequently, an
anti-neutrino can travel through an extraordinary thickness of matter
with almost zero chance of hitting stuff on its way through; the
illustration of a neutrino traveling through a light-year thick block
of lead is famous.  They, however, do interact with matter via the
weak nuclear force.  This force has a strength comparable to that of
the electromagnetic force, but the carriers of this force are very
massive, unlike photons, the massless particles that transmit the
electromagnetic force.  As a result, the weak force has a very short
range; an anti-neutrino interacts with matter only if it happens to
pass by very close to a target particle.  The chance of this happening
is extraordinarily low, and this accounts for an anti-neutrino's
ability to travel through large amount of matter.

The weak interaction between an anti-neutrino and matter can take place
in various ways.  For instance, an anti-neutrino can hit an electron that orbits an
atomic nucleus, transmitting some of its momentum to it.  Since an
electron carries an electric charge, its motion through matter is very
noticeable; a sensitive particle detector can detect the effects
produced by this motion.  For instance, an electron carrying several
million electron-volts of energy travels faster than the speed of
light in matter; this super-luminal motion creates a shock wave of
electromagnetic radiation, which is referred to as the Cherenkov
effect.  Detectors like Super-Kamiokande~\cite{superk} detect anti-neutrinos (and
neutrinos) using this effect.

Another way that an anti-neutrino can be detected is via the inverse
beta process, in which an anti-neutrino encounters a proton and comes
out transformed into a positron, while the proton is transformed into
a neutron.  This is written symbolically as follows:

\begin{equation}
\label{eqn:inv_beta}
\overline{\nu_{e}} + p \rightarrow n + e^+
\end{equation}

\noindent The symbols $\overline{\nu_e}$, $p$, $n$, and $e^+$ stand,
respectively, for anti-neutrino (electron-type), proton, neutron, and
positron.  In the study we performed, we considered detectors that use
this process for detecting antineurtrinos.  The advantages of this technique are the relatively high probability of the occurrence or the inverse beta process, and the
``double-bang" signature produced by the anti-neutrino.  That is, the
out-going particles $e^+$ and $n$ both produce signals in the
detector.  First, the $e^+$ produces a burst of light, the amount of
which is proportional to its energy (which is also closely related to
the anti-neutrino's energy).  This happens promptly after the
transformation in Eqn.~\ref{eqn:inv_beta} takes place.  The neutron,
however, rattles around for tens to hundreds of microseconds
(millionth of a second, which is a relatively long time in the present
context) until it is eventually absorbed by a proton or a dopant like
gadolinium in the target.  This absorption is followed by the emission
of gamma ray(s) of several million electron volts; a burst of light
proportional to this energy is produced.  In summary, then, an
anti-neutrino interacting in this manner produces two bursts of light
separated by a meter or so in distance, and tens to hundreds of
microseconds in time.  This double-bang signature is useful for
picking out anti-neutrino events from the large background produced,
for example, by radioactive contaminants in the detector.  The
background events produce random flashes of light, but they are not
very likely to produce the double-bang signature~\cite{rmp}.

The detection of anti-neutrinos using inverse beta decay is typically
done using a liquid scintillator detector; KamLAND is an example of
such a detector~\cite{kamland}.  Liquid scintillator is used primarily because the
amount of light produced by an anti-neutrino interaction event is very
large compared to Cherenkov radiation produced in water; greater light
yield translates to higher sensitivity (i.e. particles with lower
energy are visible) and better energy resolution.  In the present
context, however, the required detector size is of the order of one
megaton, which, for liquids, is about a cube of sides 100~m.  At this
scale, the use of liquid scintillator becomes impractical because of
the cost; water is the only economically realistic target material.
By itself, however, water cannot be used to detect the inverse beta decay
process because the second ``bang'' in the double-bang signature is
below the energy threshold.  In order to make the second bang visible,
the detector must be doped with an element such as gadolinium, which
aggressively absorbs the produced neutron and emits gamma rays above
the detector energy threshold.  Because of the very large absorption
cross section for thermal neutrons, only a 0.2\% concentration is
needed to capture 90\% of the neutrons.  For a 1~megaton detector, this
corresponds to 200 tons of the salt GdCl$_3$~\cite{gadzooks}.

The rate of detection of anti-neutrinos from a nuclear reactor in a
water detector is given by the equation below\footnote{This rate assumes 100\% efficiency above the threshold anti-neutrino energy of 1.8~MeV.  The anti-neutrino energy spectrum depends somewhat on the relative fraction of isotopes in a fission reactor; we took average values used in~\cite{kamland}, which were 0.567, 0.078, 0.297, and 0.057 for $^{235}$U, $^{238}$U, $^{239}$Pu, and $^{241}$Pu, respectively.  The rate also depends on the production cross section for the inverse beta process; the energy dependence was taken to be the same as that used in~\cite{kamland}.  Finally, it was assumed that all isotopes produce 204~MeV of thermal energy per fission; in fact, different isotopes produce somewhat different energies, but the value used here is a good-enough approximation for the present purpose.}:

\begin{eqnarray}
\label{eqn:reactor_nuebar_rate}
N & = & 3.04 \times 10^3~\mbox{events} \\
 \nonumber & &
\times \left(
\frac{T}{1~\mbox{year}} \right) \cdot \left(
\frac{M}{1~\mbox{Megaton}} \right) \\
\nonumber & &
\times \left(
\frac{P}{100~\mbox{MW}_{\mbox{\footnotesize th}}} \right) \cdot \left(
\frac{100~\mbox{km}}{D} \right)^2
\end{eqnarray}

 \noindent This equation shows that a 100~MW$_{\mbox{\footnotesize th}}$ nuclear
 reactor (about the upper limit of the power expected from a rogue
 nuclear reactor) at 100~km from a 1-Megaton detector exposed for
 1~year produces about 3000 observable events.  The number of
 anti-neutrino events from a fission bomb is given by:
 
 \begin{eqnarray}
 \label{eqn:bomb_nuebar_yield}
N & = & 2.25~\mbox{events} \\
\nonumber & & \times \left( \frac{M}{1~\mbox{Megaton}} \right)
\cdot \left( \frac{100~\mbox{km}}{D} \right)^2 \cdot \left(
\frac{Y}{1~\mbox{kton}} \right)
\end{eqnarray}

\noindent Unlike Eqn.~\ref{eqn:reactor_nuebar_rate}, which gives the
rate of anti-neutrino detection (events per year),
Eqn.~\ref{eqn:bomb_nuebar_yield} gives the total anti-neutrino yield
over 10~seconds during which most of the anti-neutrinos are released by
a fission bomb.  Based on this equation, one finds only 2.25 events
for a 1~kiloton bomb detonated at 100~km from a 1-Megaton detector.
This may seem small, but since the events arrive in a 10-second
window, the signal-to-background ratio is actually quite good.  For instance, a 2500~MW$_{\mbox{\footnotesize th}}$ reactor (typical power of a commercial reactor) 100~km away produces about 2.5 events in this time window, giving a signal-to-background ratio close to 1; at most locations, the ratio is much better than this.

Both Eqn.~\ref{eqn:reactor_nuebar_rate} and
Eqn.~\ref{eqn:bomb_nuebar_yield} are somewhat optimistic because they
were calculated assuming that the detector is sensitive to all values
of anti-neutrino energy (the inverse beta process requires at least
1.8~MeV in anti-neutrino energy).  In reality, the anti-neutrino energy
probably needs to be at least 3.8~MeV to be visible by the detector.
Only 58\% of events have energy above this.  Other data selection cuts
may decrease the event rate somewhat, probably to a total efficiency
of about 50\%.  For the sake of simplicity, we shall take the
efficiency to be 100\%.  Any result we obtain here, therefore, will be
over-optimistic by a factor of about $\sqrt{0.5}$.  In other words,
the actual sensitivity will be worse by about a factor of $\sqrt{2}
\approx 1.4$.

\section{Shielding from Cosmic Rays}

Because anti-neutrinos interact very rarely with matter, extreme care must be taken to ensure that the signal is not overwhelmed by background noise.  One way to deal with this is to increase the signal so much that it is comparable to the background.  This is what is done in short-baseline reactor detectors~\cite{rmp} and near-field reactor monitoring detectors~\cite{adam_bernstein}.  For some applications, however, the distance between the anti-neutrino source and the detector must be large.  Since the anti-neutrino flux is inversely proportional to the square of this distance, the signal rate is tiny in these situations.  To make up for this, the detector must be large and it must have a very small level of background noise.

\begin{figure*}
\includegraphics[width=50em]{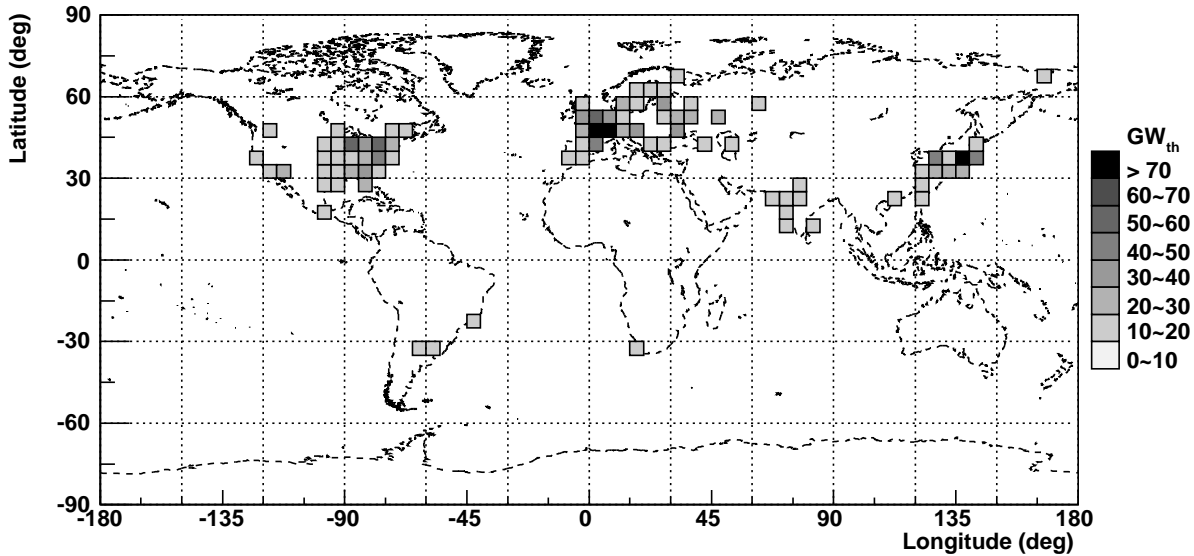}
\caption{\label{fig:world_reactor_power_map} A map of the thermal
  power of commercial and research reactors in $5^\circ \times
  5^\circ$ cells on Earth.}
\end{figure*}

There are two main classes of background noise: radioactivity present in and around the detector, and radioactivity produced by cosmic ray muons.  The level of the former can be reduced to very low levels, thanks to decades of experience from numerous neutrino detection experiments.  The latter, however, can only be reduced by brute force: i.e. by having sufficient shielding material, such as rock or water.  As a rough rule of thumb for typical anti-neutrino measurements, two kilometers of water is barely enough shielding, three kilometers is satisfactory, and four or more kilometers provides good shielding.  Shielding by rock is very costly unless it is already present, such as inside of a pre-existing mine.  Indeed, most large anti-neutrino detectors in existence today are located in commercial
mines.  For the purpose of nuclear monitoring, however, it is unlikely in general that commercial mines would exist in locations where the detectors need to be placed.  Thus, for economic reasons, the only locations where far-field monitoring detectors could be placed are in large bodies of water (i.e. oceans, seas, and large lakes).

\section{anti-neutrino Background Sources}

\begin{figure*}
\includegraphics[width=40em]{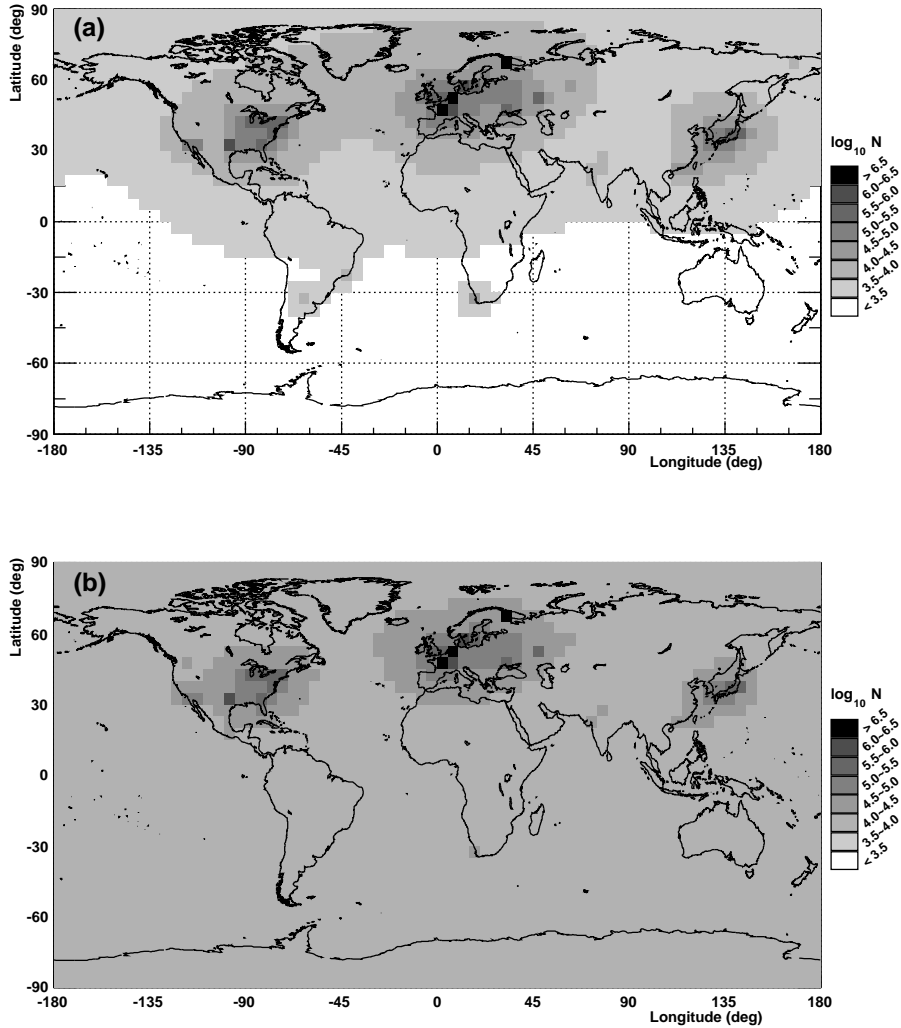}
\caption{\label{fig:world_nuebar_bkg_rate_map} The number of
  events per year (log scale) detected by a one-megaton anti-neutrino
  detector due to commercial and research reactors around the world.  (a) Assuming no georeactor.  (b) Assuming a 3~TW$_{\mbox{\footnotesize th}}$ georeactor.}
\end{figure*}

One of the most formidable background sources for nuclear monitoring
with antinuetrino detectors is the flux of anti-neutrinos from
commercial and research nuclear reactors around the world.
Distributed mostly around the northern hemisphere, a total thermal
power of about 1~TW is produced by these reactors.  The distribution
of these reactors is shown in Fig.~\ref{fig:world_reactor_power_map}.
See Appendix~\ref{app:nuc_react_loc_and_pow} for details on the
location and power of these reactors.  The number of anti-neutrinos
produced by these nuclear reactors detected per year by a one-megaton
detector located at various locations on Earth is shown in
Fig.~\ref{fig:world_nuebar_bkg_rate_map}(a).  anti-neutrinos
produced by these reactors are virtually indistinguishable from those
produced by a rogue reactor; if a rogue reactor operates in a region
where the anti-neutrino flux from commercial and research reactors is
high, it would be very difficult to detect.

Another possible source of background is the ``georeactor'', which is
a hypothetical natural nuclear reactor in the core of the Earth~\cite{h1,h2,h3,hh,h4}.  If
it exists, this reactor is expected to have a radius of several
kilometers and have a thermal power of about 1 to 10~TW$_{\mbox{\footnotesize th}}$.
Since commercial and research nuclear reactors world wide produce a total power of
about 1~TW$_{\mbox{\footnotesize th}}$, the existence of a georeactor would have a
large impact on the background rate for detecting a rogue reactor.
This is illustrated in
Fig.~\ref{fig:world_nuebar_bkg_rate_map}(b), which is the same as
Fig.~\ref{fig:world_nuebar_bkg_rate_map}(a), but with a
contribution from a 3~TW$_{\mbox{\footnotesize th}}$ georeactor.  The effect of a
georeactor is not particularly serious in much of the northern
hemisphere because the anti-neutrino flux is already high, but it
causes a serious increase in background in much of the southern
hemisphere.  For this reason, a measurement of this background is an
important prerequisite for the anti-neutrino detector array being
considered here.  A preliminary measurement has already been carried
out by KamLAND~\cite{jelena}, but the result is imprecise because of the large
background from commercial nuclear reactors.  However, a detector
whose size is comparable to KamLAND and located far away from
commercial reactors can easily make a precise measurement of
georeactor power down to about 1~TW$_{\mbox{\footnotesize th}}$.  Hanohano~\cite{geonu_conf_dec_2005} (Hawaii
anti-neutrino Observatory) is an example of a detector capable of
making this measurement.  Like the detectors in this array, Hanohano
will be placed deep in the ocean.  Thus, it is a prototype of the
megaton detectors, and the successful implementation and operation of
it would be an important prerequisite for the rogue activity detector
array concept.

%\begin{figure*}
%\includegraphics[width=50em]{../v0/world_nuebar_bkg_rate_map_3tw_gr.eps}
%\caption{\label{fig:world_nuebar_bkg_rate_map_3tw_gr} The number of
%  events per year (log scale) detected by a one-megaton anti-neutrino
%  detector due to commercial and research reactors around the world,
%  and assuming that a 3~TW$_{\mbox{\footnotesize th}}$ georeactor exists.}
%\end{figure*}

\section{Neutrino Oscillations}

A potentially important detail that must be kept in mind when considering neutrino detection is neutrino flavor oscillation.  The current view of the nature of neutrinos is that three ``flavors" of neutrinos exist; the flavors are referred to as the ``electron-type", ``muon-type", and ``tau-type".  An electron-type neutrino turns into an electron when it interacts with the target via the charged-current electroweak interaction, while a muon-type neutrino is transformed into a muon and similarly with a tau-type neutrino.  The situation with ``anti-neutrinos" -- which is the focus of this study -- is similar, except that the out-going particle has the opposite electric charge.

The importance of the above discussion is the fact that the final state particles -- the electron, muon, and the tau (and their anti-particles) -- have very different masses.  The electron, muon, and the tau have, respectively, a mass of 0.511, 105, and 1777 MeV.  A neutrino can only undergo the charged-current interaction with the target particle if it carries at least as much energy as the out-going particle mass.

Recent results of solar and reactor neutrino experiments have unequivocally established the fact that neutrinos ``oscillate".  For practical purposes, this means that the neutrino flavor when it is produced is not the same as when it is detected.  In the present context, electron anti-neutrinos are produced in a nuclear reactor; as these anti-neutrinos propagate outward, they become a quantum mechanical superposition of different neutrino flavors.  Since these anti-neutrinos have energy well below 10~MeV, that part of the superposition that has turned into a muon- or tau-type neutrino cannot interact with the target because the available energy is insufficient to produce a muon or tau.  The result is that the anti-neutrino detection rate is smaller than is expected in the absence of neutrino oscillations.  The neutrino survival probability -- defined as the fraction of detection rate compared to the rate without oscillations -- is a function of distance from the reactor.  For a threshold energy of 1.8~MeV, this starts out at 100\% for distances of 0 to several 10s of kilometers.  The probability then oscillates around an asymptotic value of 0.57 as the distance ranges from about 100~km to 300~km.  Beyond this, the amplitude of the oscillation approaches zero, and the probability is practically indistinguishable from 0.57.

In this study, we consider two cases: regional monitoring (section~\ref{sec:reg_mon}) and global monitoring (section~\ref{sec:glob_mon}).  In the former, the variation of the survival probability with distance affects the result of performance studies, so this has been taken into account in all figures and results.  In the latter, the effect of oscillation was implemented by simply scaling the no-oscillation rate by 0.57.  This simplification is valid because we are only interested in how the detector array performs as a group spanning many thousands of kilometers.  In other words, in the global scheme, we are not interested in how well several near-by detectors perform (which is covered in the regional scheme), but in how well many hundreds of widely separated detectors perform.  We have established that, for the global scheme, the asymptotic approximation of the survival probability is accurate to within a fraction of a percent.

\section{Regional Monitoring}
\label{sec:reg_mon}

\begin{figure}
\centering
\includegraphics[width=20em]{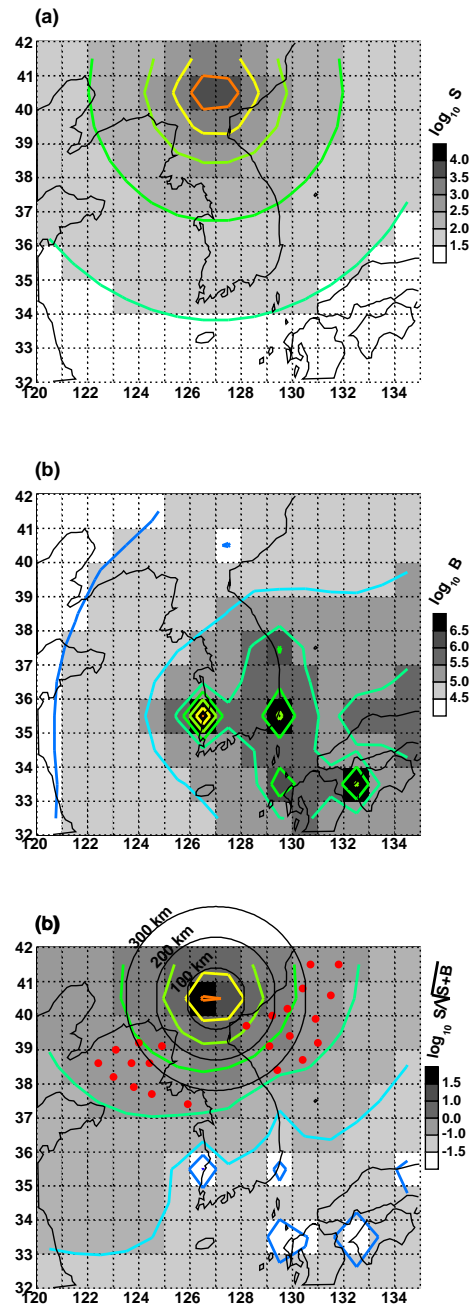}
\caption{\label{fig:sig_map_korea} The signal and background from a
  100 MW$_{\mbox{\footnotesize th}}$ rogue reactor deep in North Korean territory
  ($127.0^\circ$~E longitude, latitude $40.5^\circ$~N latitude).  (a)
  The signal $S$, defined as the number of anti-neutrino events
  detected by a 1-megaton detector exposed for one year.  (b) The
  background $B$, defined as above, but the source of anti-neutrinos
  are all commercial and research reactors around the world; the vast
  majority of detected background comes from reactors in South Korea
  and Japan. (c) The signal significance $S/\sqrt{S+B}$.  The dots
  are candidate location of the 1-megaton detectors.  Eash location was
  chosen based on the value of the significance, whose contours are
  not circular because of distortions from the background.}
\end{figure}

As an example of the capability of a megaton-scale array of
anti-neutrino detectors for the purpose of detecting rogue nuclear
activity, we consider a scenario in which rogue activity is taking
place in North Korea.  To make the illustration concrete, it was
assumed that the rogue reactor is located deep inside of North
Korean territory at longitude $127.0^\circ$~E and latitude
$40.5^\circ$~N (Fig.~\ref{fig:sig_map_korea}(a)).  North Korea
presents a realistic test case not only because of recent events, but
also because of the fact that it does not operate any nuclear reactors
legally.  If this were not the case, the monitoring regime presented
here would be easily defeated because the rogue reactor could be
placed close to a legally operated reactor, which would obscure this
activity.  Thus this monitoring regime would not be realistic for
monitoring rogue activities in South Korea or Japan, or any nation
with legally operating nuclear reactors.

\begin{figure*}
\includegraphics[width=40em]{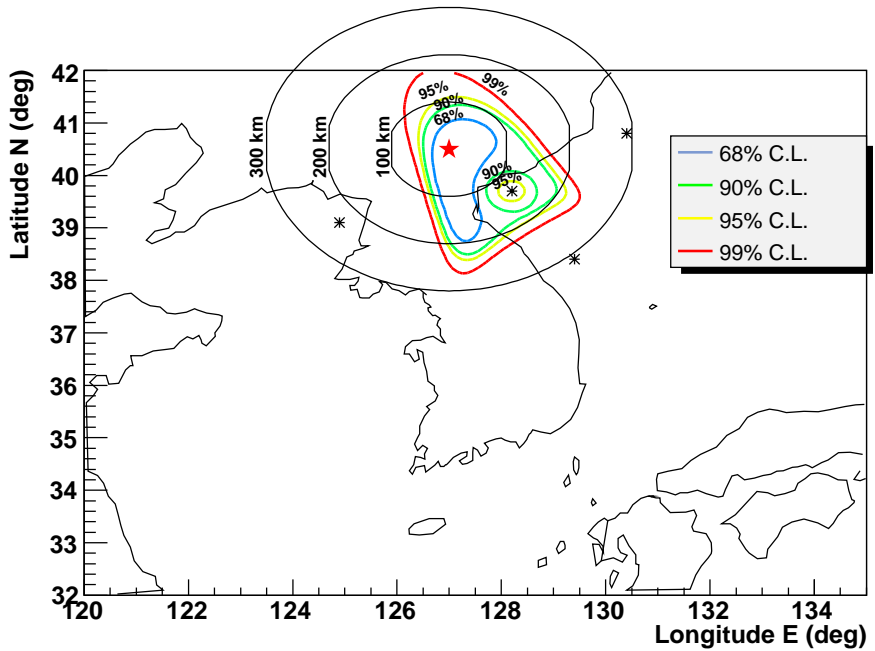}
\caption{\label{fig:nev_comm_02_05_10_21} This map demonstrates the
  ability of four 1-megaton detectors (indicated with asterisks) exposed
  for one year to detect and pin-point a 128~MW$_{\mbox{\footnotesize th}}$ rogue
  reactor (star).  The power of 128~MW$_{\mbox{\footnotesize th}}$ corresponds to the 99\% detection threshold for this configuration.  It was made by varying the position and power of the unknown reactor and comparing the number of expected events with the mean number of events that would be observed for the true reactor position; the comparison was quantified using the $\chi^2$ technique.  At each longitude and latitude, $\chi^2$ was minimized with respect to the rogue reactor power.  The contours correspond to the 68, 90, 95, and 99\% confidence level contour for two free parameters (longitude and latitude).}
\end{figure*}

The choice of location of anti-neutrino detectors should be based on
the sensitivity to rogue reactor detection.
Fig.~\ref{fig:sig_map_korea}~(a) shows the number of events deteced by
a 1-megaton detector exposed for one year, assuming the rogue reactor
power is 100~MW$_{\mbox{\footnotesize th}}$.  

Fig.~\ref{fig:sig_map_korea}~(b) shows
the number of background events, mostly from commercial nuclear
reactors in South Korea and Japan.  The sensitivity of a detector
depends on the signal $S$ and the background $B$ according to the
formula $S/\sqrt{S+B}$.  Twenty-three candidate locations were chosen
based on the sensitivity contour (Fig.~\ref{fig:sig_map_korea}~(c)).
We assumed, of course, that the detector must be located in the ocean
for cosmic ray shielding.  We did not consider the feasibility of the
candidate locations from the point of view of political boundaries,
depth, or ease of sabotage.

The general outline of the monitoring regime proceeds as follows.
First, detectors are placed in several locations around North Korea.
In our simulations, we examined array configurations with two to four
detectors, the location of which was chosen from the 23 shown in
Fig.~\ref{fig:sig_map_korea}~(c).  Of course, we do not know the rogue
reactor locaion {\it a priori}, but North Korea is not such a big
territory, so the exact choice of locations should not matter so long
as the detectors are reasonably close to land.  Second, the detectors
take data for about one year.  During this exposure period, it
receives background events from commercial reactors, but the expected
level can be calculated accurately using data provided by the reactor
operators.  Finally, one compares the observed number of events in
each detector to the expected number of background events.  If a
significant excess is observed in any of the detectors, an alarm is
raised and one would then use the data from all the detectors to try
to triangulate the location of the rogue activity.  At the same time,
political action would commence against the rogue regime.  The
statistical technique used in the comparison of the data against the
background expectation is described in detail in
Appendix~\ref{app:min_lik}.

The quantity $P_{99}$ (defined in detail in Appendix~\ref{app:min_lik}) stands
for the threshold rogue reactor thermal power that triggers an alarm
at the ``99\%'' confidence level.  Fig.~\ref{fig:nev_comm_02_05_10_21}
shows an array configuration with four detectors for which $P_{99} =
128$~MW$_{\mbox{\footnotesize th}}$.  To quantify the ability of the array to pin-point the reactor, a map of $\Delta \chi^2$ was made (Fig.~\ref{fig:nev_comm_02_05_10_21}).  This map was made by comparing the observed number of events (sum of signal and background) in each detector with the number of expected events for a hypothetical rogue reactor at different locations and power levels.  At each location, the power was varied until the $\chi^2$ between the observed and expected set of events was minimized.  As one would expect, the smallest {\it minimized} $\chi^2$ occurs at the true reactor location; $\Delta \chi^2$ is defined as the difference between this smallest minimized $\chi^2$ and that at any given location in the map (by definition, $\Delta \chi^2 = 0$ at the true location).  The contours shown in the $\Delta \chi^2$ map indicate the range of likely reconstructed positions at the specified confidence level.  In other words, an $X$-\% contour indicates that there is an $X$-\% chance that the reconstructed position would lie within the contour.

\begin{figure*}
\includegraphics[width=40em]{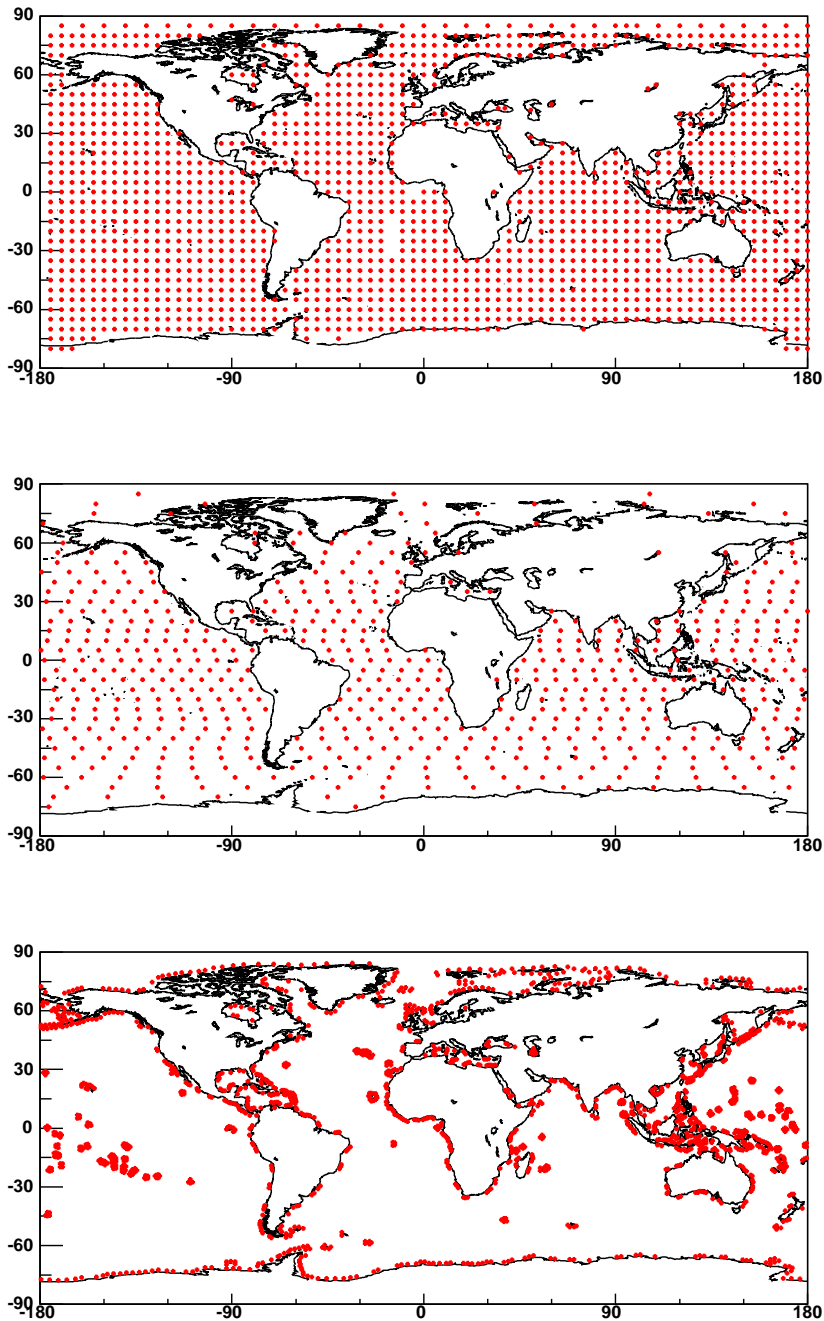}
\caption{\label{fig:world_det_distribs} The three array configurations considered in the world-wide monitoring regime.  Top: detector modules distributed on a $5^\circ \times 5^\circ$ grid in longitude and latitude.  Middle: modules distributed so that they are approximately equidistant from one another.  Bottom: modules distributed to hug coastlines; they are approximately 100~km from land, and 100~km from each other.  The number of modules in each array are 1596, 623, and 1482.  The final results were normalized so that the total detector mass is equal to 1596 modules' worth.}
\end{figure*}

\begin{figure*}
\includegraphics[width=30em]{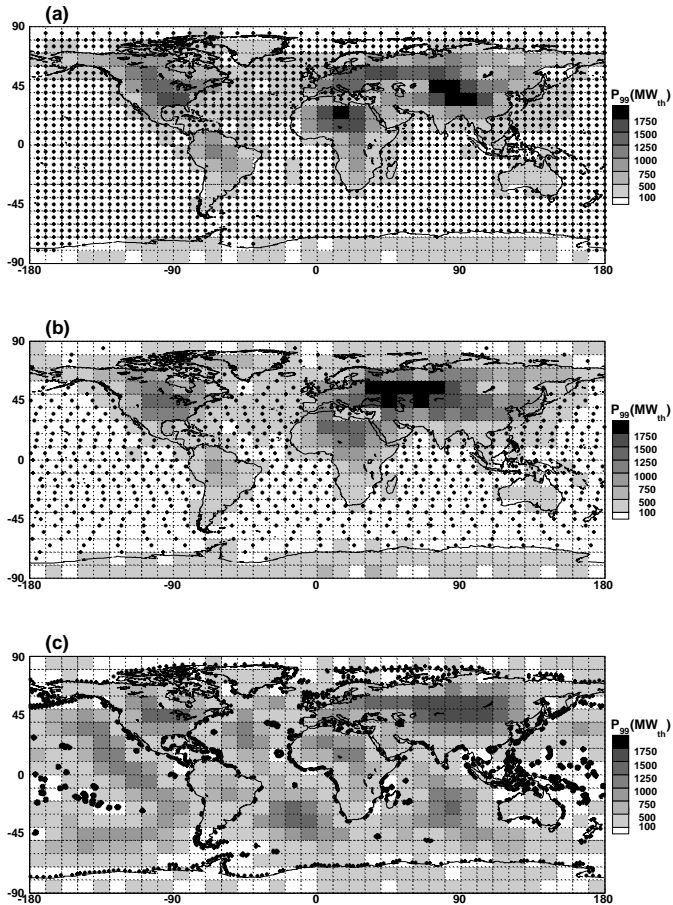}
\caption{\label{fig:world_monitor_sensitivity} Map of $P_{99}$ for the three array configurations shown in Fig.~\ref{fig:world_det_distribs}.  The power is in units of MW$_{\mbox{\footnotesize th}}$.  The number of detector modules in each array is different between arrays, but the total mass has been scaled to 1596 times 10 megatons $\approx$ 16~gigatons.  The target power of $P_{99} < 100$~ MW$_{\mbox{\footnotesize th}}$ is indicated by the white areas.}
\end{figure*}

\begin{figure*}
\includegraphics[width=0.8\textwidth]{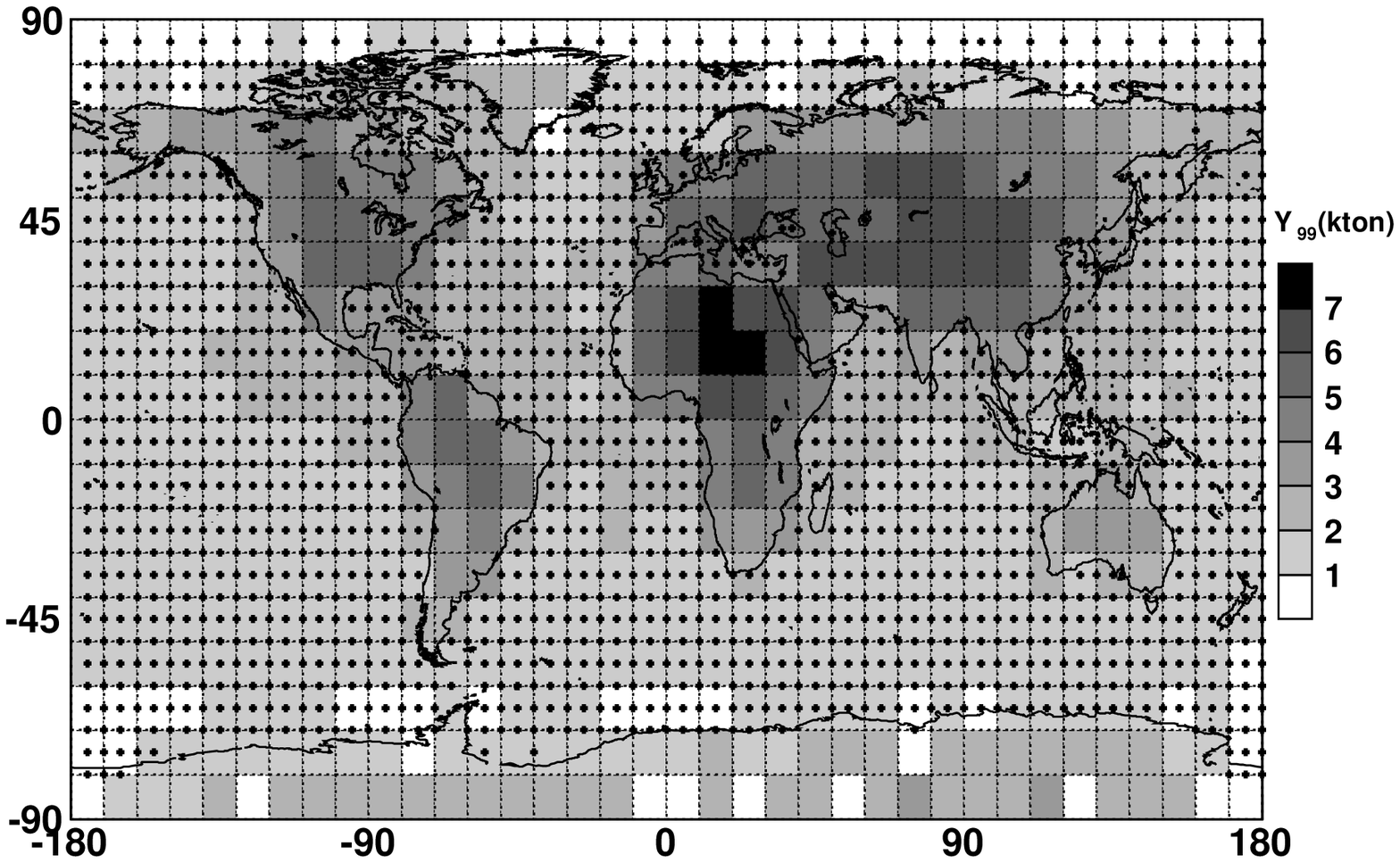}
\caption{\label{fig:world_monitor_sensitivity_bomb} Map of $Y_{99}$ for the array configuration shown in the top of Fig.~\ref{fig:world_det_distribs}.  The target sensitivity is less than 1~kton, which is indicated by the white boxes.}
\end{figure*}

There are several notable features in
Fig.~\ref{fig:nev_comm_02_05_10_21}.  First is the fact that each detector strongly rules out a circular region of radius of several 10s~km.  Second is the fact that the alarm level $P_{99}$ is determined almost completely by the
closest detector.  The addition of the
other detectors do not lower the alarm level (i.e. they do not improve
the sensitivity); their role is to help pin-point the location of the rogue reactor.  To see this, note that if only the closest detector were present, the minimum $\chi^2$ would be an annular region around it; the other detectors strongly rule out circular regions surrounding their locations, thus narrowing down the possible locations.

We finally note that nuclear reactors need to be cooled; reactors located inland are cooled with a river or a lake.  Thus the intersection of rivers and lakes with the confidence region discussed above would allow one to focus in on possible reactor sites.

\section{Global Monitoring}
\label{sec:glob_mon}

More ambitious in objective than regional monitoring is a global
monitoring regime, the goal of which is to monitor all locations on
Earth.  Unlike the regional monitoring case, one cannot optimize resources to focus in on a suspect region, so the size requirements are very demanding.  First, detector modules need to be an order of magnitude larger than in the regional monitoring case -- i.e. each module is 10~megatons, which corresponds to a cube of sides 216~m.  This is about the limiting size of a detector module from the point of view of light detection efficiency because even in extremely pure water, light has a maximum attenuation length of about 100~m~\cite{superk}.  Thus light produced in the center of the detector is attenuated by about a factor of 0.40; such events would be detected with low efficiency with any detector that is significantly larger.  Also, the number of modules in an array needs to be on the order of 1000.  As before, we took the exposure time to be one year.  We considered three different array configurations in our study, shown in Fig.~\ref{fig:world_det_distribs}.  To measure the performance of the arrays, a map of $P_{99}$ was made (see Appendix~\ref{app:min_lik}).  In other words, a rogue reactor was assumed to exist in various locations on Earth.  For each location, the rogue reactor power was varied until the reactor was detectable at the ``99\%" confidence level.  The maps show that in most costal regions (i.e. within several hundred kilometers of the shore), the array is sensitive to rogue reactor power of several hundred MW$_{\mbox{\footnotesize th}}$.  The sensitivity worsens to about 1000~MW$_{\mbox{\footnotesize th}}$ for regions with a high level of legal nuclear activity.  The sensitivity worsens yet to almost 2000 MW$_{\mbox{\footnotesize th}}$ deep within continents.  Since rogue reactors, realistically, should have a power of less than about 100~MW$_{\mbox{\footnotesize th}}$, it is seen that the world-wide monitoring regime considered here does not measure up well to the task at hand.

\section{Monitoring of Fission Bomb Detonation}

The energy production mechanism of a fission bomb is basically the same as that of a nuclear reactor.  The main difference is that the latter operate in a steady-state mode, while the former releases its energy in a short burst.  Most of the anti-neutrinos from a fission bomb are released in 10~seconds from the moment of detonation.  The anti-neutrino yield from a 1~kton bomb observed by a 1~megaton detector at 100~km from ground-zero is 2.25 events (Eqn.~\ref{eqn:bomb_nuebar_yield}).  This may seem like a small number, but one must consider the fact that the amount of background is reduced greatly by the fact that the observation time is ten seconds.  A study of the sensitivity of a global array was carried out as in the case for rogue reactors (section~\ref{sec:glob_mon}).  In this case, a map of ``$Y_{99}$" was made instead of $P_{99}$, where $Y_{99}$ is the yield of a fission bomb that can be detected at the 99\% confidence level, as defined in Appendix~\ref{app:min_lik}.  The result is shown in Fig.~\ref{fig:world_monitor_sensitivity_bomb}.  Unfortunately, at most places on Earth, the sensitivity is several kilotons, compared to the goal of one kiloton.  However, since the goal is not too far off, one may achieve the goal by targeting to certain regions (though not going down quite to the regional scale like for North Korea), or by loosening the alarm threshold.

\section{Cost}

Here, only a very rough estimate of the cost of the arrays will be attempted.  The main costs involve: (1) photomultiplier tubes (PMTs), (2) detector material, civil engineering, transport, etc., (3) water purification, (4) gadolinium dopant, and (5) man-power.  The PMT cost is fairly well-understood.  If we assume the same coverage as the Super-Kamiokande detector (about 40\% of the detector wall area), assuming that the same PMTs as Super-Kamiokande will be used, and assuming that the cost per PMT will be about \$1000 (assuming that bulk-discount or economy of scale reduces the price per PMT), about 120,000 PMT will be required per detector module.  This translates to 120 million dollars per detector.  The material, civil engineering, transport, etc. cost is not well-known at this point, but probably several hundred million dollars per module is the right order of magnitude.  The cost of water purification is also not known right now, but probably about 100 million dollars is the right order of magnitude.  Several thousand tons of GdCl$_3$ per module is required; at \$3 per kilogram, this translates to several million dollars, which is negligible compared to the total cost.  Man-power and everything else is quite vague, but something on the order of \$100 million is probably the right order of magnitude.  In total, then, each detector module will probably cost about \$1~billion.  This implies that a regional monitoring scheme would cost several billion dollars.  In contrast, a world monitoring regime will cost several trillion dollars.

\section{Conclusion}

Our study shows that targeted regional monitoring of rogue nuclear reactor activity in a nation without pre-existing legally operated nuclear reactor may be done at a cost of several billion dollars, provided that the nation has significant coast lines facing large bodies of water.  We note, however, that the cost accounting is very rough, and that several key features of the scheme have yet to be proved feasible.  For instance, the idea~\cite{gadzooks} of doping a water-based detector with gadolinium to make it sensitive to reactor anti-neutrinos is promising, but yet unproven.  The verdict should be out in the next several years as Super-Kamiokande starts its third experimental phase this year (2006).  Another unproven scheme is the deploying of a KamLAND-like detector in a deep-sea environment.  Hanohano appears to be the single experiment that will test this idea in the coming decade.  Finally, the idea of deploying a megaton-scale detector is unproven.  Ideas to construct detectors on this scale exist (e.g. Hyper-K and UNO), but there are no schedules for building any.  Moreover, these detectors are to be deployed on land, which probably simplifies matters considerably compared to deploying it deep in the ocean.  The more ambitious idea of an untargeted world-wide array was studied, but it mostly did not reach the target sensitivity of about 100~MW$_{\mbox{\footnotesize th}}$ for reactors nor the 1~kiloton yield for fission bombs.  This was at a prohibitive cost of more than trillion dollars.  Thus it is concluded that an untargeted world-wide monitoring scheme is unrealistic.  A viable monitoring scheme must focus in on some region in order to optimize resources.

%==============================================================================

\begin{acknowledgments}
I would like to thank Prof. John Learned for his leadership in initiating at the University of Hawaii a program exploring the use of anti-neutrinos for monitoring nuclear activities.  I feel honored to have been invited to join the research effort, and am indebted to him for his kind support throughout the time I was there.  I would like to thank Prof. Stephen Dye for his dedication to Hanohano and related research activities.  I would also like to thank Jelena Maricic for working with me on various aspects of research in anti-neutrino detection.  I thank all members of the University of Hawaii Hanohano team who worked hard to keep up the momentum for an exceptionally worthy goal.  Thanks also to the members of the Neutrino Physics Group and all others in the Physics Department for maintaining a high-quality research environment, and to the Department Staff whose daily support was invaluable.
\end{acknowledgments}

%==============================================================================

\appendix

\section{A List of Nuclear Reactor Location and Power}
\label{app:nuc_react_loc_and_pow}

This appendix gives a list of the location (longitude and latitude)
and the nominal thermal power of registered nuclear reactors
throughout the world.  The list was obtained in 2003 from the
International Nuclear Safety Center~\footnote{The International
Nuclear Safety Center, operated by the Argonne National Laboratory,
maintains a list of registered nuclear reactors worldwide.  Prof. John
Learned of the University of Hawaii, Manoa obtained a text version of
the list through private channels.}.  The list may not be up to date, and the positions are only approximate.
Our results, however, do not depend sensitively on the exact world wide
distribution of nuclear reactors, nor on the relatively small change
in the total nuclear power world wide since 2003, so this should be
sufficient for the purpose of this study.  A total of 433 reactors are
in the list, and the total thermal power is 1.06~TW$_{\mbox{\footnotesize th}}$.
This list will be provided in ASCII format upon request to
{\tt ehguillian@gmail.com}.

\begin{table*}
\begin{tabular}{|c|c|c|c||c|c|c|c|} \hline\hline
Reactor & Thermal & Latitude & Longitude (E) & 
Reactor & Thermal & Latitude & Longitude (E) \\
Number & Power (MW) &  &  &
Number & Power (MW) &  &  \\ \hline
   1 & 1375 &  40.166 &   44.133 &  61 & 2894 &  39.200 &   -0.633 \\
   2 & 1100 & -34.000 &  -59.250 &  62 & 3027 &  40.702 &   -2.379 \\
   3 & 2103 & -32.233 &  -64.449 &  63 &  510 &  40.350 &   -2.883 \\
   4 & 1192 &  51.316 &    4.266 &  64 &    1 &  42.766 &   -3.200 \\
   5 & 1192 &  51.316 &    4.266 &  65 & 2686 &  39.807 &   -5.696 \\
   6 & 2775 &  51.316 &    4.266 &  66 & 2696 &  39.807 &   -5.696 \\
   7 & 2988 &  51.316 &    4.266 &  67 & 2686 &  41.200 &    0.566 \\
   8 & 2660 &  50.533 &    5.266 &  68 & 2686 &  41.200 &    0.566 \\
   9 & 2775 &  50.533 &    5.266 &  69 & 2785 &  40.966 &    0.883 \\
  10 & 2988 &  50.533 &    5.266 &  70 & 2000 &  61.233 &   21.450 \\
  11 & 1375 &  43.750 &   23.633 &  71 & 2000 &  61.233 &   21.450 \\
  12 & 1375 &  43.750 &   23.633 &  72 & 1375 &  60.366 &   26.366 \\
  13 & 1375 &  43.750 &   23.633 &  73 & 1375 &  60.366 &   26.366 \\
  14 & 1375 &  43.750 &   23.633 &  74 & 2785 &  45.260 &   -0.688 \\
  15 & 1882 & -23.000 &  -44.450 &  75 & 2785 &  45.260 &   -0.688 \\
  16 & 2156 &  45.066 &  -66.450 &  76 & 2785 &  45.260 &   -0.688 \\
  17 & 2180 &  46.400 &  -72.316 &  77 & 2785 &  45.260 &   -0.688 \\
  18 & 2774 &  43.883 &  -78.750 &  78 & 3817 &  49.535 &   -1.881 \\
  19 & 2774 &  43.883 &  -78.750 &  79 & 3817 &  49.535 &   -1.881 \\
  20 & 2774 &  43.883 &  -78.750 &  80 & 2785 &  47.229 &    0.167 \\
  21 & 2774 &  43.883 &  -78.750 &  81 & 2785 &  47.229 &    0.167 \\
  22 & 1754 &  43.816 &  -79.066 &  82 & 2905 &  47.229 &    0.167 \\
  23 & 1754 &  43.816 &  -79.066 &  83 & 2905 &  47.229 &    0.167 \\
  24 & 1754 &  43.816 &  -79.066 &  84 & 3817 &  49.858 &    0.633 \\
  25 & 1754 &  43.816 &  -79.066 &  85 & 3817 &  49.858 &    0.633 \\
  26 & 2832 &  44.333 &  -81.600 &  86 & 3817 &  49.858 &    0.633 \\
  27 & 2832 &  44.333 &  -81.600 &  87 & 3817 &  49.858 &    0.633 \\
  28 & 2832 &  44.333 &  -81.600 &  88 & 4250 &  46.457 &    0.659 \\
  29 & 2832 &  44.333 &  -81.600 &  89 & 3817 &  44.106 &    0.849 \\
  30 &  947 &  46.966 &    7.266 &  90 & 3817 &  44.106 &    0.849 \\
  31 & 2806 &  47.366 &    7.966 &  91 & 3817 &  49.975 &    1.211 \\
  32 & 3012 &  47.583 &    8.149 &  92 & 3817 &  49.975 &    1.211 \\
  33 & 1130 &  47.550 &    8.216 &  93 & 2785 &  47.720 &    1.579 \\
  34 & 1130 &  47.550 &    8.216 &  94 & 2785 &  47.720 &    1.579 \\
  35 & 2905 &  22.600 &  114.533 &  95 & 2785 &  51.016 &    2.144 \\
  36 & 2905 &  22.600 &  114.533 &  96 & 2785 &  51.016 &    2.144 \\
  37 &  966 &  30.450 &  120.933 &  97 & 2785 &  51.016 &    2.144 \\
  38 & 1375 &  49.083 &   16.133 &  98 & 2785 &  51.016 &    2.144 \\
  39 & 1375 &  49.083 &   16.133 &  99 & 2785 &  51.016 &    2.144 \\
  40 & 1375 &  49.083 &   16.133 & 100 & 2785 &  51.016 &    2.144 \\
  41 & 1375 &  49.083 &   16.133 & 101 & 2785 &  47.733 &    2.516 \\
  42 & 3765 &  49.983 &   10.183 & 102 & 2785 &  47.733 &    2.516 \\
  43 & 3840 &  48.516 &   10.400 & 103 & 2785 &  47.733 &    2.516 \\
  44 & 3840 &  48.516 &   10.400 & 104 & 2785 &  47.733 &    2.516 \\
  45 & 3690 &  53.400 &   10.433 & 105 & 3817 &  47.507 &    2.877 \\
  46 & 2575 &  48.600 &   12.300 & 106 & 3817 &  47.507 &    2.877 \\
  47 & 3765 &  48.600 &   12.300 & 107 & 3817 &  48.517 &    3.520 \\
  48 & 3850 &  52.466 &    7.316 & 108 & 3817 &  48.517 &    3.520 \\
  49 & 3517 &  49.716 &    8.416 & 109 &  563 &  44.816 &    4.700 \\
  50 & 3733 &  49.716 &    8.416 & 110 & 2785 &  44.329 &    4.732 \\
  51 & 2575 &  49.250 &    8.450 & 111 & 2785 &  44.329 &    4.732 \\
  52 & 3850 &  49.250 &    8.450 & 112 & 2785 &  44.329 &    4.732 \\
  53 & 3733 &  53.433 &    8.466 & 113 & 2785 &  44.329 &    4.732 \\
  54 & 1050 &  49.366 &    9.083 & 114 & 2785 &  44.631 &    4.755 \\
  55 & 2292 &  53.916 &    9.116 & 115 & 2785 &  44.631 &    4.755 \\
  56 & 2497 &  49.033 &    9.166 & 116 & 2785 &  44.631 &    4.755 \\
  57 & 3850 &  49.033 &    9.166 & 117 & 2785 &  44.631 &    4.755 \\
  58 & 3765 &  53.850 &    9.350 & 118 & 3817 &  45.405 &    4.755 \\
  59 & 3765 &  52.033 &    9.416 & 119 & 3817 &  45.405 &    4.755 \\
  60 & 1892 &  53.616 &    9.533 & 120 & 4270 &  50.090 &    4.789 \\ \hline
\end{tabular}
\caption{\label{tab:nuc_react_list_01} A list of registered nuclear
  reactors world wide.}
\end{table*}

\begin{table*}
\begin{tabular}{|c|c|c|c||c|c|c|c|} \hline\hline
Reactor & Thermal & Latitude & Longitude (E) & 
Reactor & Thermal & Latitude & Longitude (E) \\
Number & Power (MW) &  &  &
Number & Power (MW) &  &  \\ \hline
 121 & 4270 &  50.090 &    4.789 & 181 & 3293 &  37.416 &  138.600 \\
 122 & 2785 &  45.795 &    5.270 & 182 & 3293 &  37.416 &  138.600 \\
 123 & 2785 &  45.795 &    5.270 & 183 & 3293 &  37.416 &  138.600 \\
 124 & 2785 &  45.795 &    5.270 & 184 & 3926 &  37.416 &  138.600 \\
 125 & 2785 &  45.795 &    5.270 & 185 & 3926 &  37.416 &  138.600 \\
 126 & 3817 &  49.413 &    6.216 & 186 & 1650 &  43.033 &  140.516 \\
 127 & 3817 &  49.413 &    6.216 & 187 & 1650 &  43.033 &  140.516 \\
 128 & 3817 &  49.413 &    6.216 & 188 & 3293 &  36.461 &  140.610 \\
 129 & 3817 &  49.413 &    6.216 & 189 & 1380 &  37.416 &  141.000 \\
 130 & 2660 &  47.906 &    7.565 & 190 & 2381 &  37.416 &  141.000 \\
 131 & 2660 &  47.906 &    7.565 & 191 & 2381 &  37.416 &  141.000 \\
 132 & 1375 &  46.572 &   18.854 & 192 & 2381 &  37.416 &  141.000 \\
 133 & 1375 &  46.572 &   18.854 & 193 & 2381 &  37.416 &  141.000 \\
 134 & 1375 &  46.572 &   18.854 & 194 & 3293 &  37.416 &  141.000 \\
 135 & 1375 &  46.572 &   18.854 & 195 & 3293 &  37.316 &  141.033 \\
 136 &  660 &  19.833 &   72.700 & 196 & 3293 &  37.316 &  141.033 \\
 137 &  660 &  19.833 &   72.700 & 197 & 3293 &  37.316 &  141.033 \\
 138 &  802 &  21.266 &   73.400 & 198 & 3293 &  37.316 &  141.033 \\
 139 &  801 &  21.266 &   73.400 & 199 & 1593 &  38.400 &  141.500 \\
 140 &  801 &  14.866 &   74.450 & 200 & 2436 &  38.400 &  141.500 \\
 141 &  801 &  14.866 &   74.450 & 201 & 2775 &  35.500 &  126.416 \\
 142 &  693 &  24.883 &   75.633 & 202 & 2775 &  35.500 &  126.416 \\
 143 &  694 &  24.883 &   75.633 & 203 & 2825 &  35.500 &  126.416 \\
 144 &  801 &  28.183 &   78.366 & 204 & 2825 &  35.500 &  126.416 \\
 145 &  801 &  28.183 &   78.366 & 205 & 1928 &  35.333 &  129.333 \\
 146 &  801 &  12.583 &   80.183 & 206 & 1876 &  35.333 &  129.333 \\
 147 &  801 &  12.583 &   80.183 & 207 & 2775 &  35.333 &  129.333 \\
 148 & 1650 &  33.516 &  129.833 & 208 & 2775 &  35.333 &  129.333 \\
 149 & 1650 &  33.516 &  129.833 & 209 & 2775 &  37.100 &  129.383 \\
 150 & 3423 &  33.516 &  129.833 & 210 & 2775 &  37.100 &  129.383 \\
 151 & 3423 &  33.516 &  129.833 & 211 & 2825 &  37.100 &  129.383 \\
 152 & 2660 &  31.833 &  130.200 & 212 & 2825 &  37.100 &  129.383 \\
 153 & 2660 &  31.833 &  130.200 & 213 & 2156 &  35.750 &  129.500 \\
 154 & 1650 &  33.483 &  132.316 & 214 & 2169 &  35.750 &  129.500 \\
 155 & 1650 &  33.483 &  132.316 & 215 & 1000 &  44.600 &   50.300 \\
 156 & 2660 &  33.483 &  132.316 & 216 & 4800 &  55.600 &   26.483 \\
 157 & 1380 &  35.533 &  133.000 & 217 & 4800 &  55.600 &   26.483 \\
 158 & 2436 &  35.533 &  133.000 & 218 & 1931 &  19.725 &  -96.387 \\
 159 & 2440 &  35.516 &  135.500 & 219 & 1931 &  19.725 &  -96.387 \\
 160 & 2440 &  35.516 &  135.500 & 220 & 1366 &  51.433 &    3.716 \\
 161 & 2660 &  35.516 &  135.500 & 221 &  433 &  24.866 &   66.789 \\
 162 & 2660 &  35.516 &  135.500 & 222 & 2180 &  44.316 &   28.016 \\
 163 & 3423 &  35.533 &  135.650 & 223 &   62 &  68.033 &  166.250 \\
 164 & 3423 &  35.533 &  135.650 & 224 &   62 &  68.033 &  166.250 \\
 165 & 3423 &  35.533 &  135.650 & 225 &   62 &  68.033 &  166.250 \\
 166 & 3423 &  35.533 &  135.650 & 226 &   62 &  68.033 &  166.250 \\
 167 & 1031 &  35.683 &  135.966 & 227 & 3200 &  59.500 &   29.050 \\
 168 & 1456 &  35.683 &  135.966 & 228 & 3200 &  59.500 &   29.050 \\
 169 & 2440 &  35.683 &  135.966 & 229 & 3200 &  59.500 &   29.050 \\
 170 &  714 &  35.766 &  136.000 & 230 & 3200 &  59.500 &   29.050 \\
 171 &  557 &  35.750 &  136.016 & 231 & 1375 &  67.466 &   32.466 \\
 172 & 1064 &  35.747 &  136.022 & 232 & 1375 &  67.466 &   32.466 \\
 173 & 3411 &  35.747 &  136.022 & 233 & 1375 &  67.466 &   32.466 \\
 174 & 1593 &  37.050 &  136.733 & 234 & 1375 &  67.466 &   32.466 \\
 175 & 1593 &  34.616 &  138.150 & 235 & 3200 &  54.166 &   33.233 \\
 176 & 2436 &  34.616 &  138.150 & 236 & 3200 &  54.166 &   33.233 \\
 177 & 3293 &  34.616 &  138.150 & 237 & 3200 &  54.166 &   33.233 \\
 178 & 3293 &  34.616 &  138.150 & 238 & 3000 &  57.916 &   35.083 \\
 179 & 3293 &  37.416 &  138.600 & 239 & 3000 &  57.916 &   35.083 \\
 180 & 3293 &  37.416 &  138.600 & 240 & 3200 &  51.683 &   35.616 \\ \hline
\end{tabular}
\caption{\label{tab:nuc_react_list_02} A list of registered nuclear
  reactors world wide (continued).}
\end{table*}

\begin{table*}
\begin{tabular}{|c|c|c|c||c|c|c|c|} \hline\hline
Reactor & Thermal & Latitude & Longitude (E) & 
Reactor & Thermal & Latitude & Longitude (E) \\
Number & Power (MW) &  &  &
Number & Power (MW) &  &  \\ \hline
 241 & 3200 &  51.683 &   35.616 & 301 & 1555 &  54.033 &   -2.916 \\
 242 & 3200 &  51.683 &   35.616 & 302 &  947 &  51.200 &   -3.133 \\
 243 & 3200 &  51.683 &   35.616 & 303 &  947 &  51.200 &   -3.133 \\
 244 & 1375 &  51.283 &   39.216 & 304 & 1500 &  51.200 &   -3.133 \\
 245 & 1375 &  51.283 &   39.216 & 305 & 1500 &  51.200 &   -3.133 \\
 246 & 3000 &  51.283 &   39.216 & 306 &  260 &  55.016 &   -3.216 \\
 247 & 3000 &  51.916 &   47.366 & 307 &  260 &  55.016 &   -3.216 \\
 248 & 3000 &  51.916 &   47.366 & 308 &  260 &  55.016 &   -3.216 \\
 249 & 3000 &  51.916 &   47.366 & 309 &  260 &  55.016 &   -3.216 \\
 250 & 3000 &  51.916 &   47.366 & 310 &  270 &  54.416 &   -3.483 \\
 251 &   60 &  54.233 &   49.616 & 311 &  270 &  54.416 &   -3.483 \\
 252 &  150 &  54.233 &   49.616 & 312 &  270 &  54.416 &   -3.483 \\
 253 & 1470 &  56.850 &   61.316 & 313 &  270 &  54.416 &   -3.483 \\
 254 & 2270 &  57.250 &   12.116 & 314 & 1496 &  54.416 &   -3.483 \\
 255 & 2440 &  57.250 &   12.116 & 315 & 1496 &  54.416 &   -3.483 \\
 256 & 2775 &  57.250 &   12.116 & 316 & 1760 &  53.416 &   -4.483 \\
 257 & 2775 &  57.250 &   12.116 & 317 & 1730 &  53.416 &   -4.483 \\
 258 & 1800 &  55.700 &   12.916 & 318 &  780 &  50.550 &    0.580 \\
 259 & 1375 &  57.416 &   16.666 & 319 &  780 &  50.550 &    0.580 \\
 260 & 1800 &  57.416 &   16.666 & 320 & 1550 &  50.550 &    0.580 \\
 261 & 3300 &  57.416 &   16.666 & 321 & 1550 &  50.550 &    0.580 \\
 262 & 2928 &  60.400 &   18.166 & 322 &  500 &  51.750 &    0.883 \\
 263 & 2928 &  60.400 &   18.166 & 323 &  500 &  51.750 &    0.883 \\
 264 & 3200 &  60.400 &   18.166 & 324 &  800 &  52.200 &    1.616 \\
 265 & 1876 &  45.916 &   15.533 & 325 &  800 &  52.200 &    1.616 \\
 266 & 1375 &  48.500 &   17.683 & 326 & 3411 &  52.200 &    1.616 \\
 267 & 1375 &  48.500 &   17.683 & 327 & 3817 &  33.390 & -112.862 \\
 268 & 1375 &  48.500 &   17.683 & 328 & 3817 &  33.390 & -112.862 \\
 269 & 1375 &  48.500 &   17.683 & 329 & 3817 &  33.390 & -112.862 \\
 270 & 1375 &  48.250 &   18.450 & 330 & 3390 &  33.370 & -117.557 \\
 271 & 1375 &  48.250 &   18.450 & 331 & 3390 &  33.370 & -117.557 \\
 272 & 2785 &  21.966 &  120.733 & 332 & 3323 &  46.471 & -119.333 \\
 273 & 2785 &  21.966 &  120.733 & 333 & 3338 &  35.212 & -120.854 \\
 274 & 1775 &  25.300 &  121.583 & 334 & 3411 &  35.212 & -120.854 \\
 275 & 1775 &  25.300 &  121.583 & 335 & 1998 &  41.944 &  -70.579 \\
 276 & 2894 &  25.200 &  121.666 & 336 & 3411 &  42.898 &  -70.851 \\
 277 & 2894 &  25.200 &  121.666 & 337 & 2700 &  41.309 &  -72.168 \\
 278 & 1375 &  51.333 &   25.883 & 338 & 3579 &  41.309 &  -72.168 \\
 279 & 1375 &  51.333 &   25.883 & 339 & 1593 &  42.780 &  -72.516 \\
 280 & 3000 &  51.333 &   25.883 & 340 & 3071 &  41.271 &  -73.953 \\
 281 & 3000 &  50.600 &   26.550 & 341 & 3025 &  41.271 &  -73.953 \\
 282 & 3200 &  51.383 &   30.100 & 342 & 1930 &  39.814 &  -74.206 \\
 283 & 3000 &  47.816 &   31.216 & 343 & 3411 &  39.463 &  -75.536 \\
 284 & 3000 &  47.816 &   31.216 & 344 & 3411 &  39.463 &  -75.536 \\
 285 & 3000 &  47.816 &   31.216 & 345 & 3293 &  39.468 &  -75.538 \\
 286 & 3200 &  47.483 &   34.633 & 346 & 3293 &  40.220 &  -75.590 \\
 287 & 3200 &  47.483 &   34.633 & 347 & 3293 &  40.220 &  -75.590 \\
 288 & 3200 &  47.483 &   34.633 & 348 & 3293 &  41.092 &  -76.149 \\
 289 & 3200 &  47.483 &   34.633 & 349 & 3293 &  41.092 &  -76.149 \\
 290 & 3200 &  47.483 &   34.633 & 350 & 3293 &  39.759 &  -76.269 \\
 291 & 3200 &  47.483 &   34.633 & 351 & 3293 &  39.759 &  -76.269 \\
 292 & 1500 &  54.650 &   -1.183 & 352 & 2436 &  43.524 &  -76.398 \\
 293 & 1500 &  54.650 &   -1.183 & 353 & 1850 &  43.522 &  -76.410 \\
 294 & 1555 &  55.966 &   -2.516 & 354 & 3323 &  43.522 &  -76.410 \\
 295 & 1555 &  55.966 &   -2.516 & 355 & 2700 &  38.435 &  -76.442 \\
 296 &  893 &  51.650 &   -2.566 & 356 & 2700 &  38.435 &  -76.442 \\
 297 &  893 &  51.650 &   -2.566 & 357 & 2441 &  37.166 &  -76.698 \\
 298 & 1500 &  54.033 &   -2.916 & 358 & 2441 &  37.166 &  -76.698 \\
 299 & 1500 &  54.033 &   -2.916 & 359 & 2568 &  40.153 &  -76.725 \\
 300 & 1556 &  54.033 &   -2.916 & 360 & 1520 &  43.292 &  -77.309 \\ \hline
\end{tabular}
\caption{\label{tab:nuc_react_list_03} A list of registered nuclear
  reactors world wide (continued).}
\end{table*}

\begin{table*}
\begin{tabular}{|c|c|c|c||c|c|c|c|} \hline\hline
Reactor & Thermal & Latitude & Longitude (E) & 
Reactor & Thermal & Latitude & Longitude (E) \\
Number & Power (MW) &  &  &
Number & Power (MW) &  &  \\ \hline
 361 & 2893 &  38.061 &  -77.791 & 421 & 2568 &  35.227 &  -93.231 \\
 362 & 2893 &  38.061 &  -77.791 & 422 & 2815 &  35.227 &  -93.231 \\
 363 & 2416 &  33.958 &  -78.011 & 423 & 1670 &  45.333 &  -93.848 \\
 364 & 2436 &  33.958 &  -78.011 & 424 & 2381 &  40.362 &  -95.641 \\
 365 & 2775 &  35.633 &  -78.956 & 425 & 3411 &  38.239 &  -95.689 \\
 366 & 2300 &  34.405 &  -80.159 & 426 & 3817 &  28.795 &  -96.048 \\
 367 & 2700 &  27.349 &  -80.246 & 427 & 3817 &  28.795 &  -96.048 \\
 368 & 2700 &  27.349 &  -80.246 & 428 & 1500 &  41.521 &  -96.077 \\
 369 & 2200 &  25.435 &  -80.331 & 429 & 3411 &  32.298 &  -97.785 \\
 370 & 2200 &  25.435 &  -80.331 & 430 & 3411 &  32.298 &  -97.785 \\
 371 & 2652 &  40.622 &  -80.434 & 431 & 2775 & -33.650 &   18.416 \\
 372 & 2652 &  40.622 &  -80.434 & 432 & 2775 & -33.650 &   18.416 \\
 373 & 3411 &  35.432 &  -80.948 & 433 &    0 &  40.166 &   44.133 \\
 374 & 3411 &  35.432 &  -80.948 &  &  &  &  \\
 375 & 3411 &  35.051 &  -81.069 &  &  &  &  \\
 376 & 3411 &  35.051 &  -81.069 &  &  &  &  \\
 377 & 3579 &  41.801 &  -81.143 &  &  &  &  \\
 378 & 2775 &  34.296 &  -81.320 &  &  &  &  \\
 379 & 3565 &  33.142 &  -81.765 &  &  &  &  \\
 380 & 3565 &  33.142 &  -81.765 &  &  &  &  \\
 381 & 2436 &  31.934 &  -82.344 &  &  &  &  \\
 382 & 2436 &  31.934 &  -82.344 &  &  &  &  \\
 383 & 2544 &  28.957 &  -82.699 &  &  &  &  \\
 384 & 2568 &  34.792 &  -82.899 &  &  &  &  \\
 385 & 2568 &  34.792 &  -82.899 &  &  &  &  \\
 386 & 2568 &  34.792 &  -82.899 &  &  &  &  \\
 387 & 2772 &  41.597 &  -83.086 &  &  &  &  \\
 388 & 3292 &  41.963 &  -83.259 &  &  &  &  \\
 389 & 3411 &  35.603 &  -84.790 &  &  &  &  \\
 390 & 3411 &  35.223 &  -85.088 &  &  &  &  \\
 391 & 3411 &  35.223 &  -85.088 &  &  &  &  \\
 392 & 2652 &  31.223 &  -85.112 &  &  &  &  \\
 393 & 2652 &  31.223 &  -85.112 &  &  &  &  \\
 394 & 2530 &  42.322 &  -86.315 &  &  &  &  \\
 395 & 3250 &  41.976 &  -86.566 &  &  &  &  \\
 396 & 3411 &  41.976 &  -86.566 &  &  &  &  \\
 397 & 3293 &  34.704 &  -87.119 &  &  &  &  \\
 398 & 3293 &  34.704 &  -87.119 &  &  &  &  \\
 399 & 3292 &  34.704 &  -87.119 &  &  &  &  \\
 400 & 1518 &  44.281 &  -87.536 &  &  &  &  \\
 401 & 1518 &  44.281 &  -87.536 &  &  &  &  \\
 402 & 1650 &  44.343 &  -87.536 &  &  &  &  \\
 403 & 3411 &  41.244 &  -88.229 &  &  &  &  \\
 404 & 3411 &  41.244 &  -88.229 &  &  &  &  \\
 405 & 2527 &  41.390 &  -88.271 &  &  &  &  \\
 406 & 2527 &  41.390 &  -88.271 &  &  &  &  \\
 407 & 3323 &  41.244 &  -88.671 &  &  &  &  \\
 408 & 3323 &  41.244 &  -88.671 &  &  &  &  \\
 409 & 2894 &  40.172 &  -88.834 &  &  &  &  \\
 410 & 3411 &  42.075 &  -89.282 &  &  &  &  \\
 411 & 3411 &  42.075 &  -89.282 &  &  &  &  \\
 412 & 2511 &  41.726 &  -90.310 &  &  &  &  \\
 413 & 2511 &  41.726 &  -90.310 &  &  &  &  \\
 414 & 3410 &  29.995 &  -90.471 &  &  &  &  \\
 415 & 3833 &  32.008 &  -91.048 &  &  &  &  \\
 416 & 2894 &  30.757 &  -91.332 &  &  &  &  \\
 417 & 1658 &  42.101 &  -91.777 &  &  &  &  \\
 418 & 3565 &  38.758 &  -91.782 &  &  &  &  \\
 419 & 1650 &  44.619 &  -92.633 &  &  &  &  \\
 420 & 1650 &  44.619 &  -92.633 &  &  &  &  \\ \hline
\end{tabular}
\caption{\label{tab:nuc_react_list_04} A list of registered nuclear
  reactors world wide (continued).}
\end{table*}

\section{The Statistical Technique Used to Compare the Observed Number
  of Events against the Expected Number}
\label{app:min_lik}

Let us say that there are $N_{det}$ detectors in the array.  In the
absence of any rogue reactor, detector number $i$ detects $b_i$ events
per year; these are from commercial and research reactors, and
(possibly) from the georeactor in Earth's core.  We denote the set of
observed number of events $b_1$, $b_2$, $\ldots b_{N_{det}}$ by the
usual ``set'' notation $\{ b_i \}$.

The actually observed number of events in each detector is represented
as $\{ n_i \}$.  In the absence of rogue activity, the numbers in $\{
n_i \}$ should agree with those in $\{ b_i \}$.  If, however, rogue
activity is taking place, the predicted numbers $\{ b_i \}$ is
incorrect, and it should be replaced with $\{ b_i + s_i \}$, where
$s_i$ represents the number of events in detector number $i$ due to
the rogue activity.

The method used to detect rogue activity starts with the assumption
that no rogue activity is taking place, so that the predicted number
of events at each detector is given by $\{ b_i \}$.  The set of
observed number of events is compared against the observed number $\{
n_i \}$ using a {\it likelihood function}; as the name suggests, this
function provides information about how likely a set of numbers $\{
n_i \}$ is to have resulted from the predicted set $\{ b_i \}$, given
statistical and systematic uncertainties.  For this report, we
considered only statistical uncertainty, in which case the logarithm
of the likelihood function (the log-likelihood function) is defined as
follows:

\begin{equation}
{\cal L} = -b_i + n_i \ln b_i - \ln \Gamma (n_i + 1)
\end{equation}

\noindent The last term $\ln \Gamma (n_i + 1)$ is the logarithm of the
Gamma function.

\begin{figure}
\includegraphics[width=25em]{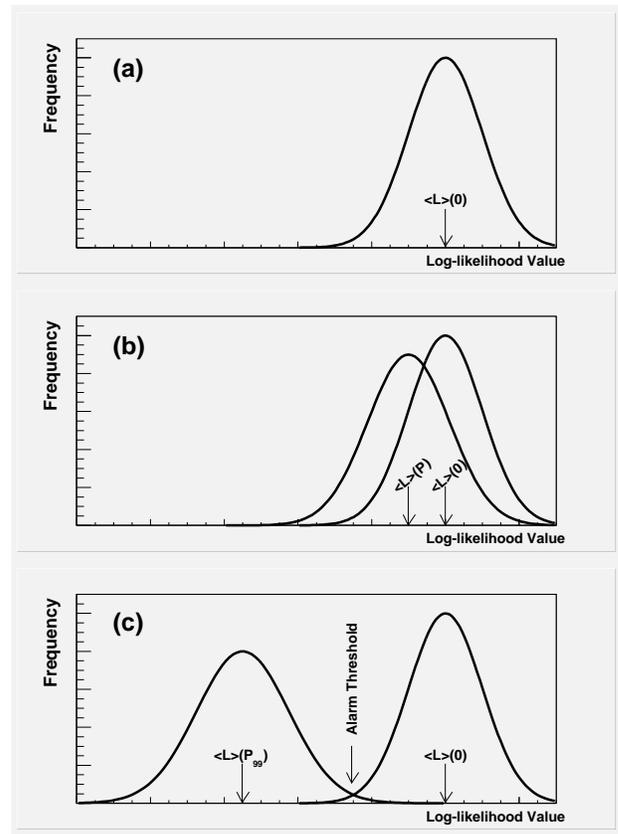}
\caption{\label{fig:ll_demo} Illustrating how the log-likelihood
  distribution changes with rogue reactor power.  (a) When no rogue
  reactor exists, the data agree well with the assumption, so that the
  mean log-likelihood value $\left< {\cal L} \right>(0)$ is high.  Any
  given measurement is distributed around the mean due to statistical
  fluctuations.  (b) As the rogue reactor power increases to $P$, the
  mean value $\left< {\cal L} \right>(P)$ decreases.  However, $P$ is
  small so the distribution at this power largely overlaps with the
  distribution at zero power, which means that the detector array is
  not sensitive enough to confidently detect the rogue reactor.  (c)
  When the rogue reactor power is sufficiently large, the overlap
  between the distributions become very small, and the existence of
  the reactor can be confirmed with great confidence.  The power
  $P_{99}$ is defined as the power above which there is 99\% chance
  that the likelihood value will be above the alarm threshold, which
  is defined as the log-likelihood value below which there is only 1\%
  chance for a ``false positive''.  We note that the width of the
  distribution increases slowly with power.}
\end{figure}

The value of ${\cal L}$ for a given measurement (lasting one year) is
not known {\it a priori} because of statistical fluctuations, although
the mean expected value $\left< {\cal L} \right>$ is
(Fig.~\ref{fig:ll_demo}~(a)).  The mean expected value, in fact,
depends on the power $P$ of the rogue reactor; we denote this
dependence as $\left< {\cal L} \right>(P)$.  As the power increases,
the assumption that no rogue reactor exists becomes increasingly
inconsistent with observations; this inconsistency causes $\left<
{\cal L} \right>$ to be biased to lower values.  When $P$ is small,
the slightly biased distribution of $\left< {\cal L} \right>(P)$
largely overlaps the distribution of $\left< {\cal L} \right>(0)$,
which implies that the detector array is not sensitive enough to
detect such a low value of $P$ (Fig.~\ref{fig:ll_demo}~(b)).  However,
as the power is raised, a point is reached where the two distributions
are different enough that the rogue reactor can be judged to exist
with great confidence (Fig.~\ref{fig:ll_demo}~(c)).

For the purpose of identifying a rogue reactor, a threshold level of
the log-likelihood value for triggering an alarm is necessary.  If
this is set too close to $\left< {\cal L} \right>(0)$, the observed
value of $\left< {\cal L} \right>$ would easily trigger a
false-positive alarm just from statistical fluctuations.  For the
purpose of the present study, we decided to tolerate a 1\%
false-positive probability (Fig.~\ref{fig:ll_demo}~(c)).  Once this
threshold is set, we can talk about the sensitivity of an array.  We
quantified this with $P_{99}$, which is the rogue reactor power that
has a 99\% chance of clearing the alarm threshold.  Since rogue
reactors are not likely to be much larger than 100~MW$_{\mbox{\footnotesize th}}$, a
promising detector array should have $P_{99}$ at this level.  Of
course, the tolerance for false-positives and -negatives chosen here
are arbitrary; looser tolerance would result in sensitivity to lower
power, but at the cost of greater chance of mis-identification and
missing an actually existing reactor.

%==============================================================================

\pagebreak
\bibliography{nuc_mon_rep_v3}% Produces the bibliography via BibTeX.

%==============================================================================

\end{document}